\documentclass[pdflatex,sn-mathphys-num,iicol]{sn-jnl}

\usepackage{lmodern}
\usepackage{anyfontsize}
\usepackage{graphicx}%
\usepackage[inkscapelatex=false]{svg}
\usepackage{multirow}%
\usepackage{amsmath,amssymb,amsfonts}%
\usepackage{amsthm}%
\usepackage{mathrsfs}%
\usepackage[title]{appendix}%
\usepackage{xcolor}%
\usepackage{textcomp}%
\usepackage{manyfoot}%
\usepackage{booktabs}%
\usepackage{algorithm}%
\usepackage{algorithmicx}%
\usepackage{algpseudocode}%
\usepackage{listings}%
\usepackage{adjustbox}
\usepackage{graphicx}
\usepackage{textcomp} 
\usepackage{geometry} 
\usepackage{lineno}

\begin{document}

\title[]{Large spin accumulation signals in ultrafast magneto-optical experiments.}

\author[1]{\fnm{Alberto} \sur{Anadón}}
\equalcont{These authors contributed equally to this work.}
\author[1]{\fnm{Harjinder} \sur{Singh}}
\equalcont{These authors contributed equally to this work.}
\author[1]{\fnm{Eva} \sur{Díaz}}
\author[1]{\fnm{Yann} \sur{Le-Guen}}
\author[1]{\fnm{Julius} \sur{Hohlfeld}}
\author[2]{\fnm{Richard B.} \sur{Wilson}}
\author[1]{\fnm{Gregory } \sur{Malinowski}}
\author[1]{\fnm{Michel} \sur{Hehn}}
\author*[1]{\fnm{Jon} \sur{Gorchon}}\email{jon.gorchon@univ-lorraine.fr}

\affil[1]{\orgname{Université de Lorraine, CNRS, IJL}, \city{Nancy}, \postcode{F-54000}, \country{France}}
\affil[2]{\orgname{Department of Mechanical Engineering and Materials Science and Engineering Program, University of California}, \city{Riverside, CA}, \country{USA}}

\abstract{Magneto-optical techniques have become essential tools in spintronics, enabling the investigation of spin dynamics in the ultrafast regime. A key challenge in this field has been to accurately isolate the contributions to magneto-optical signals of spin transport phenomena from the local magnetization dynamics. The contribution of transported and accumulated spins was long believed to be orders of magnitude smaller than that of the magnetization and thus previous approaches to disentangle these signals have relied on specific experimental designs, usually including thick metal layers. Here, we present experimental evidence demonstrating that the magneto-optical signal from ultrafast spin accumulations can, under certain conditions, be comparable to or even exceed that of the magnetic layer in a standard ultrafast demagnetization experiment. Our findings provide a new framework for accessing and isolating these spin accumulations, allowing for time and depth dependent probing of transported spin and/or orbital angular momentum.}

\maketitle


\section*{Introduction}\label{sec:intro}

Historically, magneto-optical probes have played a vital role in unveiling and exploring various spintronic phenomena \cite{Kato2004,MihaiMiron2010, beach_dynamics_2005,yamanouchi_current-induced_2004,Liu2018,Stamm2017a,ChoiNatureTi}. Furthermore, in combination with femtosecond laser sources, magneto-optics offers a direct window into ultrafast spin dynamics \cite{beaurepaire1996ultrafast,Igarashi2023,Malinowski2008ControlMomentum}. Typical experiments consist of a first intense femtosecond-wide laser pulse which excites (i.e. pumps) a magnetic system, and a second lower-fluence one which probes the changes in the magnetic system via the magneto-optical Kerr effect (MOKE). During excitation electrons absorb the optical energy, which then gets redistributed locally among other electrons, phonons and spins, and non-locally as it is transported into neighboring layers. This redistribution of energy also results in a redistribution of angular momentum, and may result in the generation of ultrafast spin currents. In fact, such spin currents are nowadays paramount for the understanding of all-optical magnetization switching \cite{Radu2011,Igarashi2023} and broadband spintronic THz emitters \cite{Seifert2016}.




While advances on alternate probes such as THz emission \cite{Beaurepaire2004,Rouzegar2022}, ultrafast X-ray based methods \cite{Pfau2012,Boeglin2010}, or time-resolved photoemission \cite{Lisowski2005} have undoubtedly confirmed the loss of magnetization and presence of THz spin currents, MOKE remains the most widely used method to characterize ultrafast magnetization dynamics. However, the role of Kerr rotation $\theta_k$ and ellipticity $\epsilon_k$ and its connection with the magnetic properties of thin films in the ultrafast regime is still somewhat controversial. In 2000, Koopmans et al. found a significant difference between $\theta_k$ and $\epsilon_k$ in the first picoseconds after the excitation of a Ni thin film grown on a Cu(111) substrate \cite{Koopmans2000}, ascribing the differences to non-magnetic optical contributions. In contrast, Guidoni et al. soon after showed that in a CoPt$_3$ film grown on sapphire the $\theta_k$ and $\epsilon_k$ had the same time evolution \cite{Guidoni2002}. Many more works tried to test the validity of MOKE at the time, both experimentally and theoretically, with no clear consensus \cite{Kampfrath2002Fe,BKoopmans_2003,vanKampenPhysRevLett.88.227201,Zhang2009ParadigmMagnetism,Carva2011IsSolved}. Noting the use of conductive Cu substrates \cite{Koopmans2000} vs sapphire \cite{Guidoni2002}, some works have since argued that the observed differences are likely related to spin transport and have attempted to probe and model the depth profile of the magnetization by carefully comparing $\theta_k$ and $\epsilon_k$ signals \cite{Wieczorek2015,Razdolski2017}. Other works have attempted to probe directly the resulting spin accumulation directly via linear and non-linear MOKE on thick and opaque non-magnetic layers \cite{Choi2014thermal,Choi2014a,Choi2015,Melnikov2011,Ortiz2022,choi_magneto-optical_2018}. By using such thick layers, the authors could avoid the probe from reaching the magnetic layer. Notably, Hofhferr et al.\cite{Hofherr2017} attempted to disentangle the spin accumulation and demagnetization MOKE signals in a Ni/Au bilayer by shining the probe directly on the ferromagnetic layer. This was achieved by cleverly mixing the $\theta_k$ and $\epsilon_k$ with a quarter wave-plate and suppressing the MOKE contribution from the Ni layer. Unfortunately, likely due to the limited signal-to-noise ratio, this experiment has not been reproduced yet. In fact, to this day, spin accumulation signals have never been considered as an important contribution to MOKE in usual TR-MOKE experiments.

Here we demonstrate that, under certain common conditions, spin accumulation signals can be as large or even larger than the magneto-optical signals attributed to the magnetic layer, leading to possibly confusing interpretations. By studying a series of wedged magnetic/non-magnetic bilayer systems we are able to directly probe the pure demagnetization, spin accumulation or a mixed signal, and deconvolute each contribution. We also show that differences between MOKE rotation and ellipticity may arise from the magneto-optical signal due to spin accumulation. In the case of Cu, we show that ellipticity is a much better probe of the demagnetization dynamics in the ferromagnet, whereas the rotation signal contains an important sensitivity to spin accumulation. Importantly, we demonstrate that the spin accumulation signal can also be sizable in experiments with low repetition rate amplified systems, opening the door for spin accumulation detection with most pulsed laser systems and up to extreme fluences. Strikingly, we are able to generate and measure spin accumulations up to half a mrad, on the order of full magneto-optical signals of common ferromagnets, despite the low spin-orbit coupling in Cu. All our measurements are reasonably well fitted by an ultrafast spin-diffusion model, which includes magneto-optical sensitivities based on a transmission matrix model. Our results open the door to new ways to detect spin (or orbital) angular momentum in common systems and should improve the interpretation of MOKE signals.


\begin{figure*}[ht]
\centering
                \centerline{\includegraphics[width=0.6\textwidth]{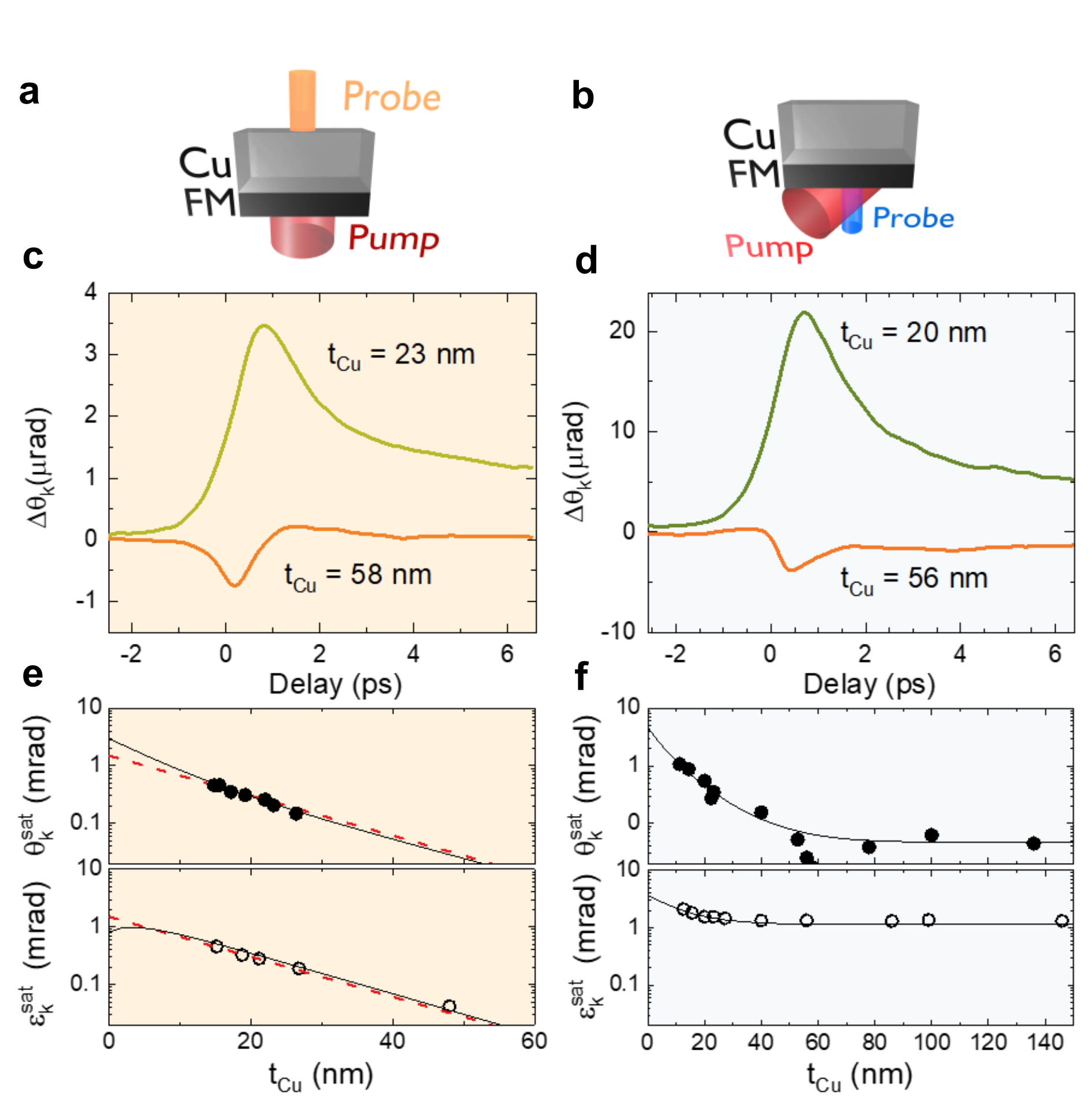}}
                \caption{\textbf{Experimental configuration, as well as time-resolved and static magneto-optical Kerr responses in the CoNi/Cu structures.} \textbf{a,b} shows Schematic for the two different experimental configurations where the pump is always incident on the FM-side (through the sapphire substrate), whereas the probe is incident either on the \textbf{a} Cu-side or \textbf{b} the FM-side. Time-resolved magneto-optical measurements with the pump beam always directed towards the FM side and the probe beam is directed to the Cu \textbf{c} and CoNi \textbf{d} sides for two different Cu thickness. We see that the peak in both cases changes sign and shows at a lower delay time for the thicker Cu. \textbf{e} and \textbf{f} panels show the rotation and ellipticity at saturation (no incident pump) from the Cu side \textbf{e} and the FM side \textbf{f} as a function of the Cu thickness. The dashed red lines in \textbf{e} are an exponential decay fit to the data, while the continuous black likes are a fits using a transmission-matrix-method simulation\cite{suppmat}.}
                \label{fig:MOKEDC}
\end{figure*}

\section*{Results}\label{sec:results}

\textbf{Setup and samples:} We fabricated three wedged samples of sapphire//Ta(3)/Cu(5)/ [Ni(0.7)/Co(0.2)]{$_4$}Cu($t_{Cu}$)/Al(3), where the thickness appears in parenthesis, expressed in nanometers. The Cu layer of the three samples has a variable thickness $t_\text{Cu}$ ranging from 10 to 30 nm, 20 to 60 nm and 50 to 150 nm, for the three samples (Methods). All Co/Ni multilayers have a perpendicular magnetic anisotropy. The top Al layer should be naturally passivated and was added to protect the Cu from oxidation. For simplicity, from here on, we will refer to the ferromagnetic Ta/Cu/[Co/Ni] section of the stack as the \textit{FM-side}, and to the top Cu/Al as the \textit{Cu-side}. Most experiments were carried with an 80 MHz Ti-Sa femtosecond laser system which provides a few nanojoules of energy (Methods). Alternatively, a 5 KHz amplified Ti-Sa system was also used (Methods).

\textbf{Dynamic magneto-optics:} We first performed pump-probe time-resolved magneto-optical Kerr effect (TR-MOKE) experiments with the 80 MHz laser system. Experiments were performed by exciting (i.e. pumping) the FM side through the transparent substrate, and probing on both the sides of the sample at normal incidence, as depicted in Figs.\textbf{\ref{fig:MOKEDC}a-b}. For easy identification of the experiment configuration, when probing the Cu side we identify the graphs by an orange background shade, whereas when probing through the FM side the shade is blue. In Figs.\textbf{\ref{fig:MOKEDC}c} and \textbf{d}, we plot the change in Kerr rotation induced by the pump beam $\Delta \theta_k$ as a function of pump-probe delay.

As shown in Figs.\textbf{\ref{fig:MOKEDC}c}, t$_{Cu}$=20 nm is thin enough that the probe can still see the magnetic layer through the Cu layer, and we observe a typical demagnetization trace\cite{beaurepaire1996ultrafast}, showing a first rapid drop in magnetization and a subsequent slower recovery as the magnetic system cools down. However, when the thickness is increased to t$_{Cu}$=58 nm, light reaching the magnetic layer is drastically reduced and therefore, as previously shown\cite{Choi2014}, the magneto-optical signal can safely be attributed to the spin accumulation resulting from the ultrafast spin currents generated during demagnetization (see \cite{suppmat} for measurements up to 200 nm in Cu thickness).

When we probe through the FM-side, as seen in Fig. \textbf{\ref{fig:MOKEDC}d} for t$_{Cu}$=23 nm, we also observe a typical demagnetization trace. However, for t$_{Cu}$=56 nm, we can see that the trace changes significantly, showing a peak with different timing, opposite polarity and lower amplitude. In this configuration, the Cu layer does not prevent the probe from reaching the magnetic layer, making these changes  unexpected.

\textbf{Static magneto-optics:} We thus investigated the static magneto-optical signals of the multilayer by probing the saturation Kerr rotation $\theta_k^{sat}$ and ellipticity $\varepsilon_k^{sat}$ as a function of t$_{Cu}$, from both the Cu and FM sides. As shown in Fig.\textbf{\ref{fig:MOKEDC}e}, when probing the Cu-side, both $\theta_k^{sat}$ and $\varepsilon_k^{sat}$ follow a Beer-Lambert-like exponential decay (red dashed line) given by $e^{-4\pi\lambda^{-1} {\rm k_{Cu}} t_{Cu}}$, where ${\rm k_{Cu}}$ is the imaginary part of the optical index of Cu, and $\lambda$ is the optical wavelength in vacuum. Since $\theta_k^{sat}$ is divided by 10 when t$_{Cu}$ is changed from 23 nm to 58 nm, the orange curve in Fig.\textbf{\ref{fig:MOKEDC}c} stills contains a very small contribution of the ferromagnet's magneto-optical signal. When probing from the FM-side, we observe that $\theta_k^{sat}$ is reduced by over an order of magnitude as t$_{Cu}$ increases past 50 nm. This is not the case for $\varepsilon_k^{sat}$, where a smaller decay is observed, but the value remaining at around 1 mrad.

\textbf{Magneto-optical model:} To model the magneto-optical signals we used a generalized transmission-matrix-model with non-diagonal susceptibility terms \cite{suppmat,10.1063/1.368058,PhysRevB.43.6423}  determined by the complex magneto-optical Voigt constants \textbf{Q}. All diagonal (non-magnetic) susceptibility terms where fixed from literature values, and the magnetic layer's \textbf{Q$_{Co/Ni}$} was fitted resulting in the solid lines shown in panels Figs.\textbf{\ref{fig:MOKEDC}e,f}. We obtained a value of \textbf{Q$_{Co/Ni}$}=0.012-0.016$i$, in good agreement with values from Ref.\cite{atkinson_fundamental_1996}. The extremely good fit allows us to predict the evolution of the magneto-optical signal as a function of thickness with high accuracy. The multiple reflections and interferences within the structure when the Cu thickness is high, are directly responsible for the drastic reduction of the rotation signal.

\begin{figure*}[ht]
\centering
                \includegraphics[width=0.7\textwidth]{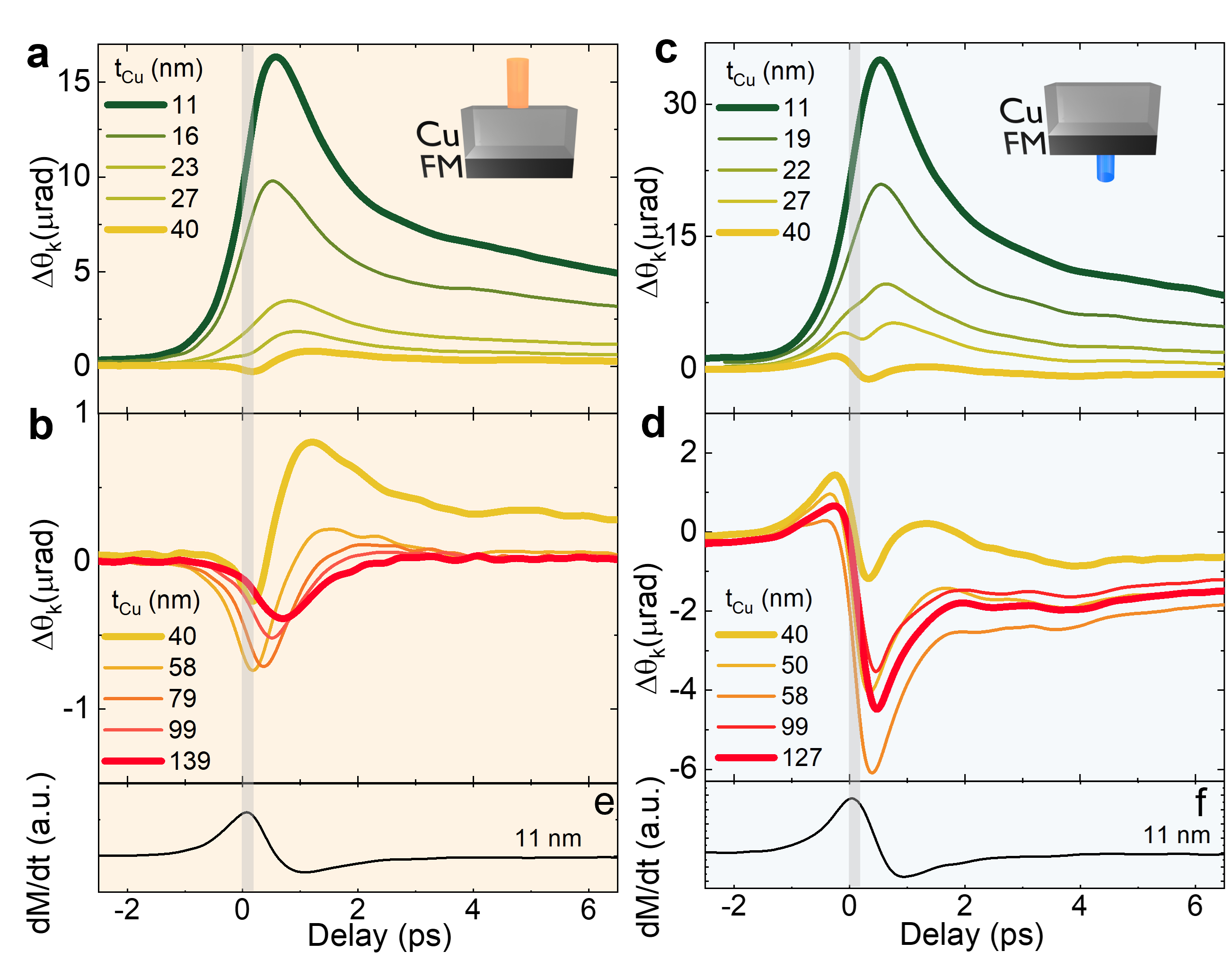}
                \caption{\textbf{Mixed demagnetization and spin accumulation signals in CoNi/Cu(t).} TR-MOKE signal for varying Cu thickness with the probe beam incident on the \textbf{a,c} CoNi layer side and the \textbf{b,d} Cu layer side. At lower t$_{Cu}$, the curves for both configurations resemble a classical demagnetization curve. At medium thickness, an additional peak appears, pointing downward in both cases. At higher t$_{Cu}$, this peak clearly dominates over the demagnetization peak. At high Cu thickness, when the probe hits the Cu side, a mostly symmetric peak is observed, and $\Delta\theta_{k}$ returns to zero after a few picoseconds. In the configuration where the probe hits the CoNi layer, the peak is clearly asymmetric and a flat background is observed after the peak. Panels \textbf{e} and \textbf{f} show the derivative with respect to the delay time of the curve with t$_{Cu}$ = 11 nm for both probe configurations. The grey shaded areas represent the peaks of the curves in panels \textbf{e} and \textbf{f}. }
                \label{fig:TR-MOKE}
\end{figure*}

\textbf{Full thickness dependence:} The full t$_{Cu}$ dependence of $\Delta \theta_k$ for both Cu-side and FM-side configurations is depicted in Figs. \textbf{\ref{fig:TR-MOKE}a-d}. In Fig.\textbf{\ref{fig:TR-MOKE}a} we show that as t$_{Cu}$ increases from 11 to 27 nm, the peak $\Delta \theta_k$ is reduced due to the reduction in magneto-optical sensitivity of the FM (see Fig.\textbf{\ref{fig:MOKEDC}e}). At t$_{Cu}$ = 40 nm, the trace becomes bipolar due to the emergence of a new earlier negative peak, which we associate with spin accumulation. Fig.\textbf{\ref{fig:TR-MOKE}b} shows a closeup of the 40 nm trace, and its evolution as t$_{Cu}$ increases up to 139 nm. For higher t$_{Cu}$, the magneto-optical signal from the FM is drastically reduced, and only a pure spin accumulation signal remains, in agreement with previous works \cite{Choi2014,Choi2014thermal,Choi2014a,Choi2015,choi_magneto-optical_2018}. The decreasing amplitude and increasing delay time of the negative peak is related to the diffusive transport of the angular momentum in the Cu film, from which an average velocity of 0.17 nm/fs can be extracted\cite{suppmat} (a fraction of the usual 1 nm/fs Fermi velocity).

Fig.\textbf{\ref{fig:TR-MOKE}c} shows the $\Delta \theta_k$ from 11 to 40 nm when probing the FM-side. Similarly, we see a reduction of the peak $\Delta \theta_k$ as the magneto-optical sensibility to the FM is reduced with the increase of t$_{Cu}$ (see Fig.\textbf{\ref{fig:MOKEDC}f}). Around t$_{Cu}$ = 27 nm we start to see a change in the shape of the trace, with a negative peak forming, which we \textit{also} attribute to spin accumulation. At thicker t$_{Cu}$ (Fig.\textbf{\ref{fig:TR-MOKE}d}) this negative signal increases, reaching a maximum at t$_{Cu}\approx$ 60 nm. To our surprise, despite the probe impinging the FM-side, the increase in t$_{Cu}$ and reduction in $\theta_k^{sat}$ results in a larger sensitivity to the spin accumulation \textit{behind} the FM, rather than to the magnetization of the FM itself. Interestingly, we believe the long-lived tail in Fig.\textbf{\ref{fig:TR-MOKE}d} is likely linked to the spin-dependent Seebeck effect\cite{Choi2015}. We observed similar but smaller long-lived signal when probing through the Cu-side as studied in \cite{Choi2015}. Importantly, most literature works assume that many ps after optical excitation, artifacts in $\theta_k$ signals can be safely excluded. Our data clearly shows this is not always true, and therefore care should be taken when, for example, using a magnet as a transducer for temperature estimations\cite{kimling_thermal_2017,Choi2015,Uchida2008,Slachter2010ThermallyMetal}.

Finally, in the lower panels \textbf{e} and \textbf{f} of Fig.\ref{fig:TR-MOKE}, we show the derivative with respect to the delay time for the curve with t$_{Cu}$ = 11 nm (dM/dt) shown in panels \textbf{a} and \textbf{b}, respectively, which follow the demagnetization. Based on prevailing theories for spin-current generation by ultrafast demagnetization, the peak of the dM/dt should correspond with the maximum of the spin current generation\cite{Choi2015}, which we identify by a gray shaded area. As it can be seen, spin accumulation peaks for relatively thin layers (ex: t$_{Cu}$=58 nm) almost coincide with the dM/dt peaks\cite{sup_note_dmdtpeak}.

\begin{figure*}[ht]
\centering
                \includegraphics[width=0.7\textwidth]{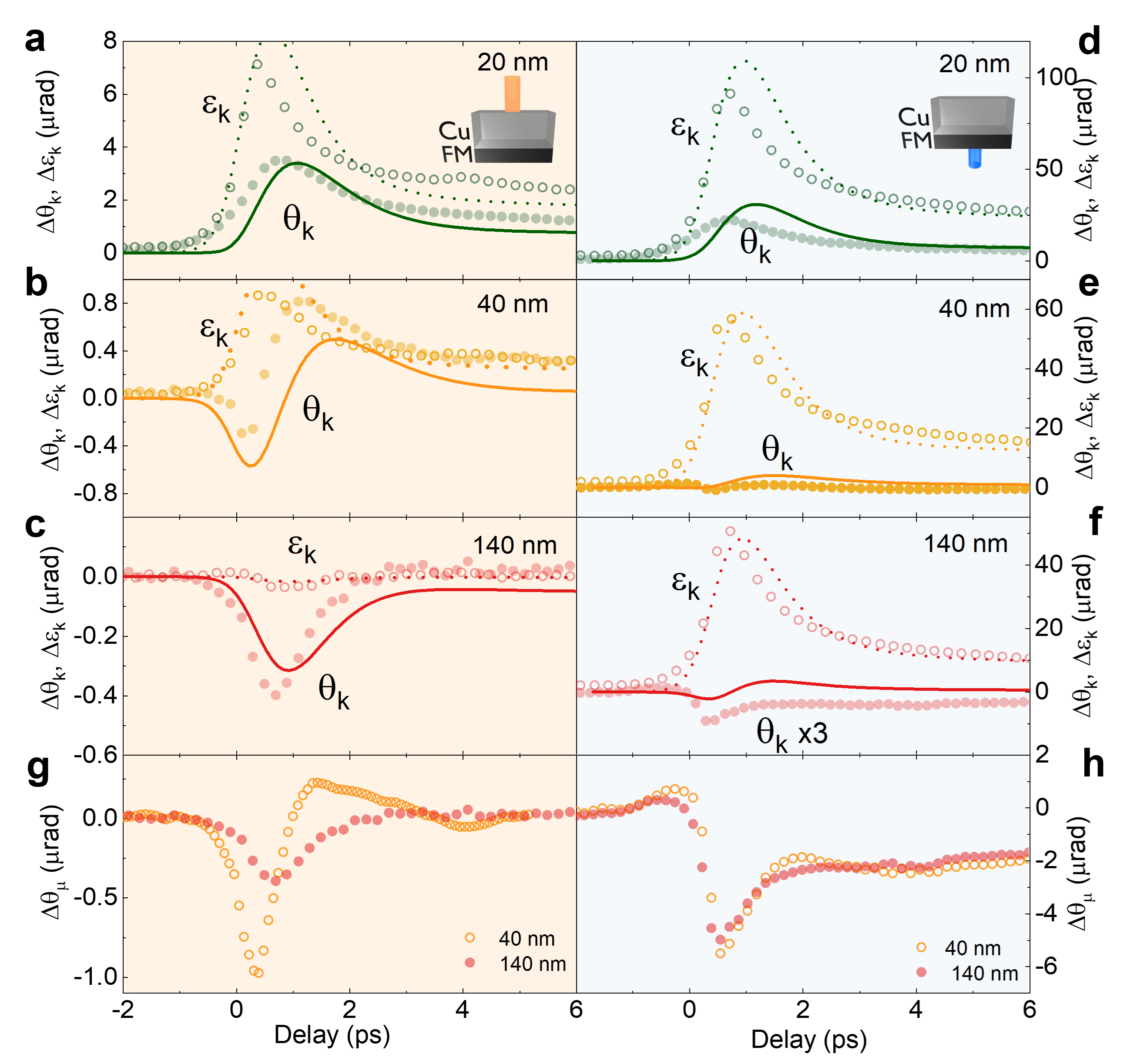}
                \caption{\textbf{Time-resolved magneto-optical signal for rotation and ellipticity and extracted spin accumulation contribution.} \textbf{a-f}. TR-MOKE rotation (filled circles) and ellipticity (open circles) signals for t$_{Cu}\approx$ 20, 40, and 140 nm, for both Cu-side (\textbf{a,b,c}) and the FM-side (\textbf{d,e,f}) probing configurations. Solid and dashed lines are simulations\cite{suppmat} with a single set of magnetic, thermal and optical parameters, except for the fluence which was modified between Cu-side and FM-side experiments.\textbf{g-h}. Extracted spin accumulation contribution to the magneto-optical signal ($\Delta \theta_\mu$) from the ellipticity and rotation signals (see text for details).}

                \label{fig:rot-ellip}
\end{figure*}

\begin{figure}[ht]
\centering
                \includegraphics[width=0.33\textwidth]{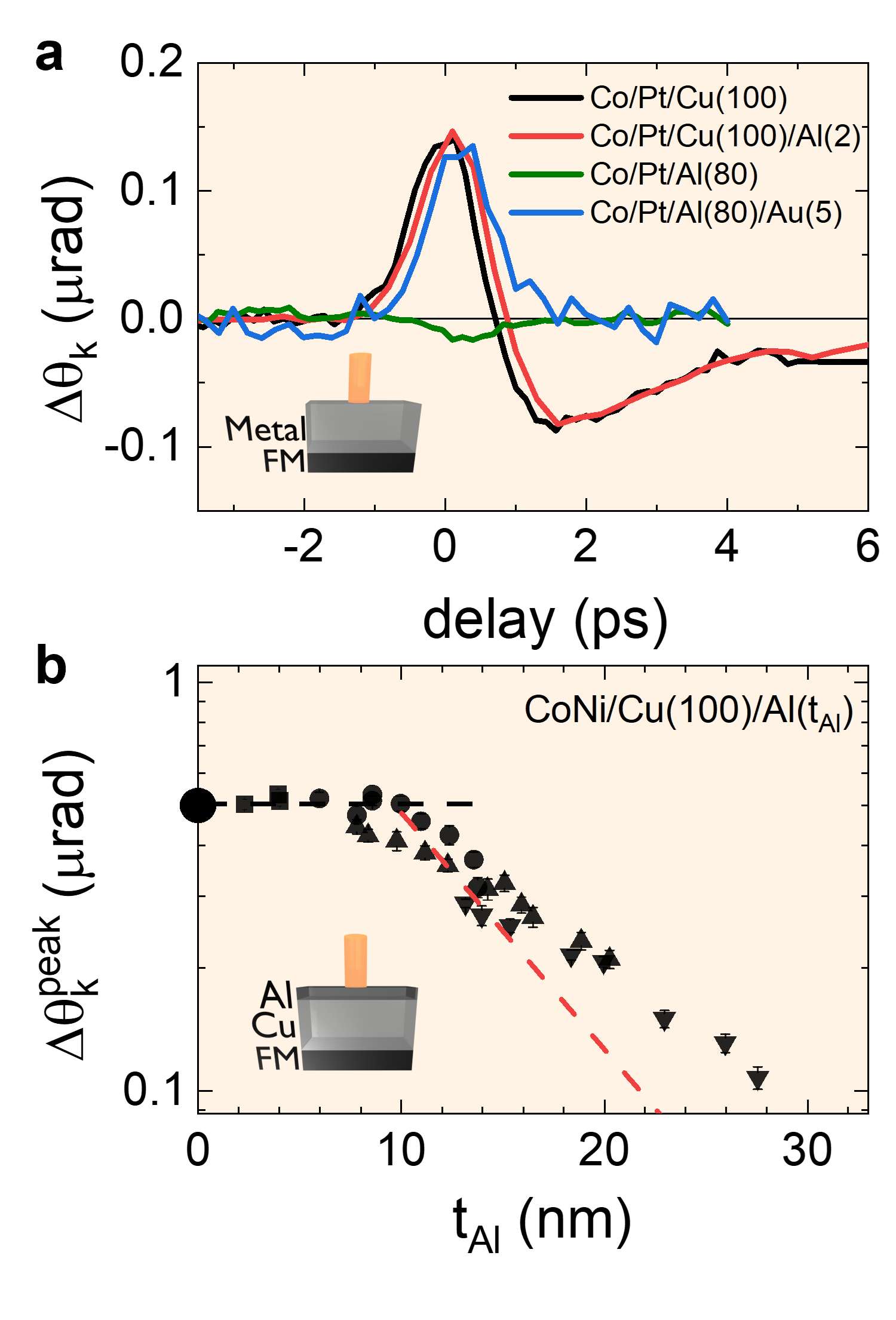}
                \caption{\textbf{Impact of different capping layers in the spin accumulation signal.} \textbf{a} TR-MOKE signal from the Cu side with different capping layers. The common part of the stacks is sapphire//Ta(3)/Pt(10)/[Co(0,82)/Pt(1)]$_2$/Co(0,82) abbreviated as Co/Pt. The different samples are , Co/Pt//Cu(100)/Al(2.5), Co/Pt//Al(80) and Co/Pt//Al(80)/Au(5). \textbf{b} Scaling of the spin accumulation peak with the thickness of an Al capping layer measured on various wedges covering the range from 0 to 30 nm. The red dashed line corresponds to the Beer-Lambert law $e^{-4\pi\lambda^{-1} {\rm k_{Al}} t_{Al}}$, with ${\rm k_{Al}=8.5}$ being Al's optical index imaginary part \cite{Ordal:88}.}
                \label{fig:Al}
\end{figure}

\begin{figure}[ht]
\centering
                \includegraphics[width=0.33\textwidth]{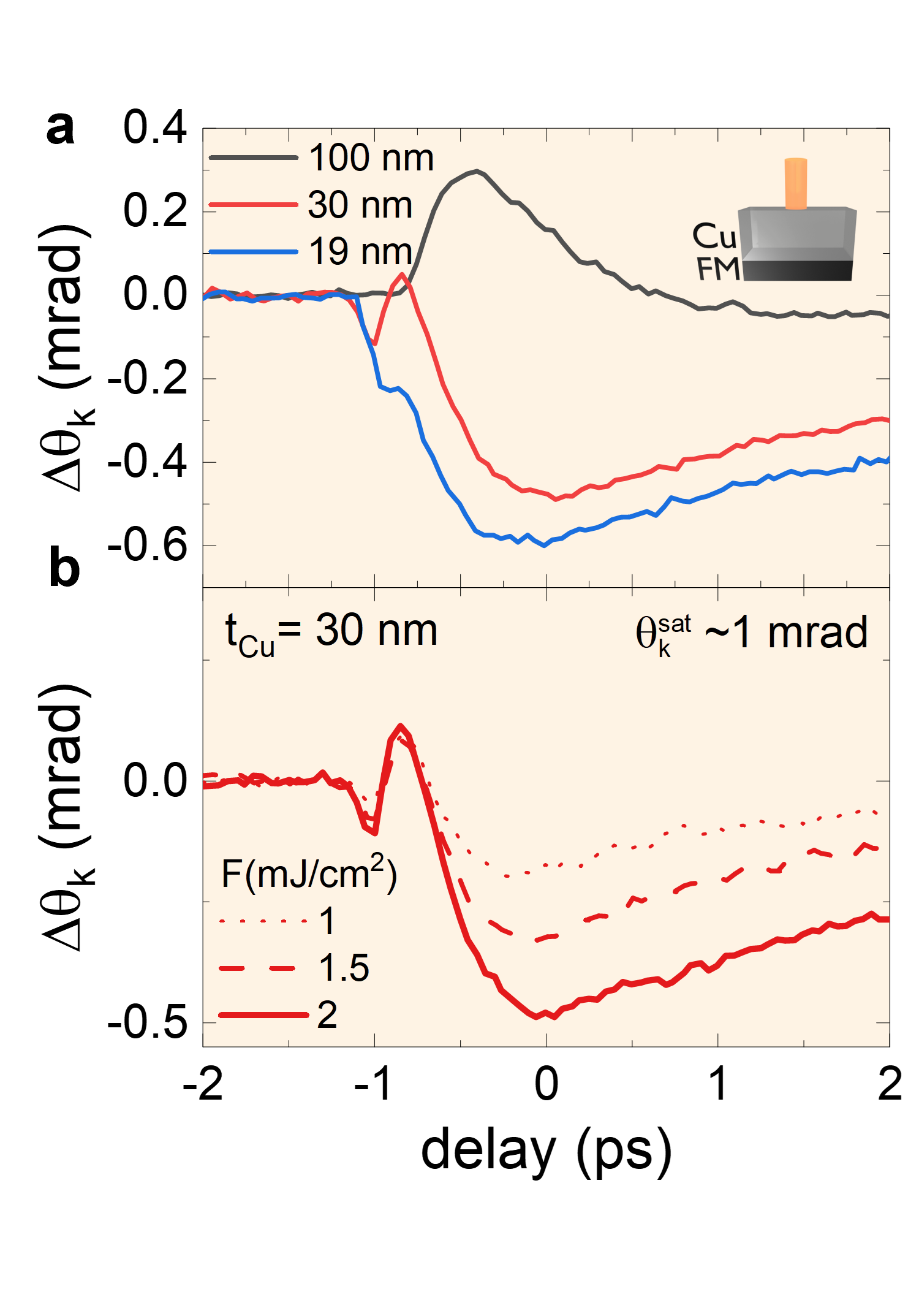}
               \caption{\textbf{High peak power time-resolved magneto-optical signal.} Dependence of the TR-MOKE signal in CoNi/Cu(100) using the amplified laser system (see Methods) for \textbf{a} different Cu thicknesses at a constant fluence of 2 mJ/cm$^2$ and for \textbf{b} different fluences at constant thickness of t$_{Cu}$=30 nm. A similar behavior is observed in the TR-MOKE signal when compared to the low-energy laser system as seen in Figure \ref{fig:TR-MOKE} at lower fluence, showing a demagnetization-dominated curve at lower t$_{Cu}$ and a spin accumulation-dominated curve at high t$_{Cu}$, while a mixed state is observed at t$_{Cu}$=30 nm. As we increase the fluence, we observe that the spin accumulation peak increases along with the demagnetization.}

                \label{fig:Harjinder}
\end{figure}

\textbf{Rotation vs ellipticity:} In Fig.\textbf{\ref{fig:rot-ellip}} we show the comparison between  $\Delta 
 \theta_k$ and $\Delta \epsilon_k$ signals for a few selected thicknesses. On the one hand, $\Delta\theta_k$ follows the previously described trends, probing either the magnetization (Figs.\textbf{\ref{fig:rot-ellip}a,d}), the spin accumulation (Figs.\textbf{\ref{fig:rot-ellip}c,f}) or a mixture of both (Figs.\textbf{\ref{fig:rot-ellip}b,e}). $\Delta \epsilon_k$ on the other hand, generally shows a demagnetization like trace (Figs.\textbf{\ref{fig:rot-ellip}a,b,d,e,f}) with a decaying amplitude as t$_{Cu}$ is increased\cite{note_bump}. At t$_{Cu}$= 140 nm (Figs.\textbf{\ref{fig:rot-ellip}c}) the sensitivity to the magnetic layer is orders of magnitude lower, so that we could attribute any sizeable signal to the spin accumulation in the Cu layer. From it, we can determine the sensitivity of $\epsilon_k$ to spin accumulation in the Cu, which is drastically lower than that of $\theta_k$ (in agreement with Ref.\cite{choi_magneto-optical_2018,uba_magneto-optical_2017}). Therefore, in order to measure the actual magnetization dynamics one has to be careful with the sensitivities to spin accumulation, and in the case of Cu, a measurement of $\epsilon_k$ should give the best results when using a wavelength of around 800 nm.

\textbf{Multi-physics model:} In order to verify the origin of these observed signals we developed a model\cite{suppmat} to estimate the ultrafast demagnetization in the low fluence limit, predict time-dependent spin density profiles and extract the resulting magneto-optical signals. In order to estimate the demagnetization, we first coupled an optical absorption transmission matrix model \cite{10.1063/1.1325200} and 3 temperature model\cite{beaurepaire1996ultrafast} to predict phonon, electron and spin temperatures. By assuming a phenomenological (spin) temperature dependence of the magnetization, we could calculate the evolution of the magnetization. We then used a simple spin diffusion model, where the source term is proportional to $dM/dt$, the change in magnetization. Additionally a small spin-dependent Seebeck effect was added, as described in Ref.\cite{choi_thermal_2015-12}. Finally, by setting a Voigt constant for the Cu layer (based on Ref.\cite{uba_magneto-optical_2017}, see also Ref.\cite{suppmat}), we calculate for each time-step the resulting $\Delta\theta_k$ and $\Delta\epsilon_k$, which we convolute with the probe's temporal shape. In the end, for a given set of parameters, we are able to fit all 12 curves in Figs.\textbf{\ref{fig:rot-ellip}} corresponding to various samples and probing directions, shown as solid (for $\Delta\theta_k$) or dashed (for $\Delta\epsilon_k$) lines. $\Delta\epsilon_k$ is mostly sensitive to the magnetization dynamics, and results in rather good fits, giving confidence in the estimated $dM/dt$. $\Delta\theta_k$, which is also sensitive to spin accumulation, is reasonably well fitted too. The relative amplitudes of the signals and the main features such as polarities of the curves are well described, but not all details are captured by the model. Given the large number of layers and parameters, it is difficult to quantify the injected spin density, but with the used parameters we estimate peaks spin densities at the top Cu surface of around 5 A/m. Finally, by assuming that the magneto-optical signal from Co/Ni and Cu layers is additive, we can extract pure spin accumulation Kerr signal $\Delta \theta_\mu$ traces from the experimental data by the following operation $\Delta \theta_\mu = \theta_k^{sat}(\Delta\theta_k/\theta_k^{sat} - \Delta\epsilon_k/\epsilon_k^{sat}) $ and as plotted in Figs.\ref{fig:rot-ellip}.g-h. This method allows one to extract spin accumulations for thinner Cu layers, even when signals are somewhat mixed.

---

\textbf{Discussion:} At this point, one might wonder about the origin of the Kerr signal.  We first turn our attention to the top Cu/air interface, and investigate its role by modifying the capping. Here we use Co/Pt as the ferromagnet, which results in a rather bipolar spin accumulation signal, as previously shown \cite{Choi2014a}. We first compare the signal from a bare 100 nm Cu layer (self-passivated) with one capped by 2.5 nm of (self-passivated) Al, but observe no differences (black vs red in Fig.\textbf{\ref{fig:Al}}.a). Pt and Ta cappings resulted\cite{suppmat} in a reduction of $\Delta \theta_k$, as reported in \cite{choi_magneto-optical_2018}.  We then verify the possible signal due to the presence of metallic Al (green in Fig.\textbf{\ref{fig:Al}}.a) but observed a complete loss of the signal\cite{Choi2014a}. As a last test, shown in Fig.\textbf{\ref{fig:Al}}.b, we increased the thickness of the Al capping, from 2.5 nm to 30 nm, to make sure the Cu layer was not oxidized at all. Indeed, from atomic force microscopy measurements\cite{suppmat} we know that the thick Cu has a roughness of $~$3 nm for 150 nm Cu films, which could potentially enable oxidation of the Cu even with the thin capping. As the cap thickness is increased, initially no change in the signal is observed, up to around 8 nm, where the signal starts to drop. We attribute the reduction in signal to the fact that the Al layer self-passivates up to a certain thickness (potentially around 8 nm on our very rough layers) and the rest of the aluminum stays metallic. Therefore, the metallic Al layer, which doesn't contribute to MOKE rotation, blocks the probe from reaching the sensitive Cu layer (red fit corresponds to $e^{-4\pi\lambda^{-1} {\rm k_{Al}} t_{Al}}$, with ${\rm k_{Al}=8.5}$ being Al's optical index imaginary part \cite{Ordal:88}. However, this fit over-estimates the loss in signal, which means the metallic Al likely enhances somehow the resulting magneto-optical signal. Nevertheless, it appears that the oxidation state of the Cu interface doesn't play a big role in the magneto-optical detection of spin-accumulation. This result suggests a bulk-sensitivity. This interpretation is strongly backed by our simulations, which only include bulk sensitivity and fit qualitatively well the data in Fig.\ref{fig:rot-ellip}.


We note that, interestingly, adding a few nanometers of Au on top of the thick Al(80) layer recovers the Kerr signal (blue in Fig.\textbf{\ref{fig:Al}}.a), even though the dynamics are different to those in Cu, probably due to the differences in optical absorption and thermal transport between Cu and Al\cite{choi_note}.

Finally, even-though we have only considered spin in our discussion and analysis, one may wonder if the reported signals could be of orbital origin instead\cite{ChoiNatureTi,Seifert2023}. Moreover, magneto-optics are particularly sensitive to orbital angular momentum, without requiring spin-orbit coupling, unlike for sensing spin \cite{ChoiNatureTi}. We do believe that, if present, orbital currents may be addressable and distinguishable from spin by these optical methods, by designing the right multilayer and experimental protocol. However, this goes beyond the scope of the present work.

\textbf{High-fluence experiments:} Given the importance of ultrafast spin transport for technological applications, we decided to extend our studies to higher fluence regimes, accessible with most amplified laser systems, and necessary for high intensity THz emission\cite{Seifert2016} and all-optical switching\cite{Igarashi2023,Radu2011}. To this aim, we used an amplified Ti:Sa laser (methods) to induce large spin dynamics on the same Co/Ni films and detect $\theta_k$. In Fig.\textbf{\ref{fig:Harjinder}.a} we show $\Delta \theta_k$ for various Cu thicknesses when probing the Cu-side and pumping the FM-side with 2 mJ/cm$^2$. As in previous low intensity experiments, we see a transition from a pure demagnetization-like trace, to a mixed one, and to a full spin-accumulation one, for t$_{Cu}\approx$ 19, 30 and 100 nm respectively (for a discussion on delays see Methods). In Fig.\textbf{\ref{fig:Harjinder}.b} we show experiments at a fixed thickness of 30 nm, but with increasing fluence, reaching demagnetizations of around 20, 35 and 50$\%$ for 1, 1.5 and 2 mJ/cm$^2$ respectively. Both the demagnetization trace and the spin accumulation increase proportionally to the absorbed fluence, but because the baseline of the spin accumulation peak is being reduced (due to the demagnetization signal), the peak appears at the same position for all fluences. When pushed to the limit, for nearly full demagnetization conditions and for a thickness of around 50 nm (where spin accumulation signals are the largest), we can reach spin accumulation signals on the order of 0.5 mrad. Such $\theta_k$ are comparable to the full $\Delta \theta_k$ obtained from thin ferromagnets (see Fig.\textbf{\ref{fig:MOKEDC}.e-f} or Refs \cite{Choi2014thermal,jhuria_spinorbit_2020} and even thick Fe films\cite{ZAK1990107}. Even if this value is not directly comparable, due to different Voigt constants in each case, it is remarkable that a non magnetic film can be polarized to such an extent by its magnetic Co/Ni neighbor.


\section*{Conclusion}\label{sec:conclusion}

Our work demonstrates that ultrafast spin accumulations can lead to large magneto-optical signals, at times comparable or larger than the magneto-optical signal originating from magnetic layers. We show that for the case of Cu at a wavelength of 785 nm, spin accumulation induces mainly a Kerr rotation and negligible ellipticity, which can be leveraged to extract the pure demagnetization and spin accumulation temporal profiles for any Cu thickness, even when probed through the ferromagnetic layer. Finally, we show that amplified systems can also be used for these kind of experiments, resulting in up to $>$0.5 mrad Kerr rotation signals. Our conclusions are backed by simulations combining ultrafast heat and spin-diffusion with a magneto-optical transmission matrix method. This work shows that with proper care, time-resolved magneto-optical experiments become a unique tool to offer a direct view into the ultrafast spin (and/or orbital) transport. These developments may prove crucial for the understanding and development of all-optical switching in spin-valves\cite{igarashi_optically_2023} or THz emission in thin multilayers\cite{Seifert2016,Seifert2023}.


\backmatter

\section*{Methods}\label{sec11}

\textbf{Sample growth}. Samples were prepared using magnetron sputtering in an Ar atmosphere of  4x10$^{-3}$ mbar and with a base pressure less then 2x10$^{-8}$ mbar. In this study double sided polished sapphire substrates were used for depositing the thin films. The primary FM stack used in this study is Ta(3)Cu(5)[Ni(0.7)/Co(0.2)]{$_4$} and referred to as Co/Ni. The values inside parentheses are thicknesses in nm. Three different wedge samples were fabricated with the magnetic stack of Co/Ni: Co/Ni/Cu(\textit{t})/Al(3) by varying the Cu thickness within the following ranges: 10 to 30 nm (wedge-1), 20 to 60 nm (wedge-2), and 50 to 150 nm (wedge-3). All the samples have perpendicular magnetic anisotropy (PMA). 
To protect the topmost Cu layer from the oxidation the samples have thin protective layer of Al that passivates in contact with the air. (See supplementary for more details about wedge samples). Moreover, besides Co/Ni, we deposited other magnetic stack: Ta(3)/Pt(10)/[Co(0,82)/Pt(1)]{$_2$}/Co(0,82) referred to as Co/Pt. 

To study the effect of the capping layer on spin accumulation, we deposited samples both with and without an Al capping layer: Co/Pt/Cu(100)/Al(2.5) and Co/Pt/Cu(100) respectively. Additionally, a wedge sample using Co/Ni as FM: Co/Ni/Cu(100)/Al(\textit{t}) by varying Al thickness ranges from 2 to 30 nm was also fabricated.

In addition to this, we have also studied spin accumulation in materials other than Cu. For this we have deposited samples with Co/Pt as FM but Al as NM with the following stack: Co/Pt/Al(80) and Co/Pt/Al(80)/Au(5), Co/Pt/Cu(100)/Pt(5), Co/Pt/Cu(100)/Ta(5).

\textbf{Experimental setups.} Two different Ti:sapphire femtosecond lasers were used. 

The main laser system is a Chameleon Vision-S\textsuperscript{\tiny\textregistered} (from Coherent\textsuperscript{\tiny\textregistered}) oscillator with a repetition rate of 80 MHz, a 785 nm center wavelength and a 12 nm bandwidth. We separate pump and probe beams by using very sharp edge-filters (Semrock\textsuperscript{\tiny\textregistered} SP01-785RU and LP02-785RE) which split the spectrum and avoid possible interference effects between pump and probe. The pump pulse-duration after spectrum filtering was measured at around 100 fs whereas the probe is at around 600 fs. The pump has a gaussian profile with a full width half maximum of around 50$\pm$2 $\mu$m. When the probe is incident at 45\textdegree (Cu-side experiments), the full-width half maximum is estimated at ~50 $\mu$m. The probe is focused through a long-working distance objective resulting in a size of around 5$\mu$m. We used for all experiments a pump power of 180mW, resulting in an incident fluence of around 0.1 mJ/cm$^2$.  We used a quarter-wave plate to change the probe sensitivity from rotation to ellipticity (see Suppl. Mat.\cite{suppmat} S5-6). For determining the saturation rotation and ellipticities, we use a chopper to modulate the probe (for lock-in detection), block the pump and perform hysteresis cycles. The values $\theta_k^{sat}$ and $\epsilon_k^{sat}$ correspond to the half difference between traces obtained for positive and negative saturation fields. For pump probe experiments we modulate the pump at 1.1 MHz with an electro-optic modulator. Reported $\Delta \theta_k$ and $\Delta \epsilon_k$ correspond to the half difference for measurements with opposite saturated states. For more details on the signal analysis (see Suppl. Mat.\cite{suppmat} Sec 5-6).

The second setup, used for data shown in Fig.\ref{fig:Harjinder}, is a Legend\textsuperscript{\tiny\textregistered} regenerative amplified laser (from Coherent\textsuperscript{\tiny\textregistered}) with a repetition rate of 5 KHz and pulse width of 25 fs. The pump and probe wavelengths are centered at 800 nm and 400 nm respectively, by using a BBO crystal to double the probe. The full beam width at 1/${e^2}$ of the pump is 243$\pm$9 $\mu$m whereas for the probe it is around 50$\mu$m. The pump is incident at normal incidence, whereas the probe is incident at a few degrees away from normal incidence.

\textbf{Magnetic field during experiments.} In all experiments, an out of plane magnetic field is  applied only to saturate the ferromagnetic layer and is turned off during the measurement. In Suppl. Mat. S9) we show possible artifacts in some configurations due to presence of the magnetic field during scans. The strength of the applied field is larger than the coercive field of the magnetic layer. Presented experimental data always corresponds to the half difference between traces obtained for positive and negative saturation fields. In a few instances, at very high powers (inset of Fig.5b), the field is maintained constant during experiments to reset the magnetization between pulses.

\textbf{Zero time delay accuracy.} We note that the zero delay between Cu-side and FM-side experiments is different due to the differences in the optical paths used, and has been arbitrarily defined. For FM-side experiments, the sample position does not affect the zero delay, as both pump and probe travel in the same direction and will experience the same shift in delay. However, in back-front type experiments, for the Cu-side experiments, any shift in sample position affects the arrival time of both the pump and probe beams in an opposite way. Therefore a shift of $\Delta x$ in sample position, will result in a shift in $2\Delta x / c$ in zero delay, c being the speed of light. On the one hand, on the 80 MHz setup, benefiting from the shallow depth-of-field of the objective, the accuracy on the sample position is around 5 $\mu$m, resulting in an uncertainty of around 30 fs. On the other hand, with the 5 kHz setup, we estimate the positional uncertainty to be around 200 $\mu$m, resulting in large shifts reaching the picosecond . For this reason, temporal shifts should not be compared on Fig.\ref{fig:Harjinder} when the sample is changed.

\section*{Acknowledgements}
This work was supported by the France 2030 government investment plan managed by the French National Research Agency under grant references PEPR SPIN – TOAST ANR-22-EXSP-0003 and SPINMAT ANR-22-EXSP-0007. This work was also supported by the French PIA project “Lorraine Université d’Excellence”, reference ANR-15-IDEX-04-LUE. Alberto Anadón acknowledges the project 3D-Sky (GAP-101108063), part of the Marie Skłodowska-Curie Actions from the Horizon Europe Program of the EU. Views and opinions expressed are those of the author(s) only and do not necessarily reflect those of the European Union. Neither the European Union nor the granting authority can be held responsible for them.
\section*{Declarations}

\begin{itemize}
\item Competing interests: The authors declare no competing interests.
\item Availability of data: Data are available upon reasonable request to the corresponding author.
\item Code availability: The authors
will agree to share the code upon reasonable request. 
\end{itemize}

\section*{Contributions}
J.G. designed the experiments and supervised the study. A.A. and H.S. performed the magneto-optical experiments. A.A., E.D., H.S. G.M. and J.G. built and optimized the optical setups. M.H. performed the wedged sample deposition and optimized the magnetic properties. A.A. made the ruler patterns by UV-lithography, and Y.L.G. performed the AFM characterizations. A.A., H.S., G.M. and J.G. analyzed the data. J.G. build the magneto-optical code, and integrated the heat and spin diffusion modules as written by R.B.W.. J.G. performed all the numerical simulations with inputs from Y.L.G.. A.A. and J.G. wrote the manuscript with input from all the authors.

\noindent


\bibliography{sn-bibliography,references,references1}

\begin{thebibliography}{52}
\bibitem[\protect\citeauthoryear{}{}]{sup_note_dmdtpeak}
\begin{botherref}
In some other similar samples a perfect coincidence was found between the dm/dt peak and the spin-accumulation peak when probing the FM-side. We have not yet understood the origin of the delay or lack of it yet between these samples.
\end{botherref}
\endbibitem
\bibitem[\protect\citeauthoryear{}{}]{note_bump}
\begin{botherref}
{We note a little bump in the data at around 4 ps, best visible on the 20 nm sample, which we attribute to a small back reflection of the pump on the sapphire/air interface (delay corresponds to a return trip through the substrate) which induces a second demagnetization.}
\end{botherref}
\endbibitem
\end{thebibliography}

\end{document}